\documentclass[11pt,cite]{article}
\usepackage{latexsym}

\usepackage[latin1]{inputenc}

\usepackage{graphicx}
\usepackage{color,pst-plot}

\usepackage{amsmath}
\usepackage {amsfonts,amssymb,amstext}

\newcommand{\lbl}[1]{\label{eq:#1}}
\newcommand{ \rf}[1]{(\ref{eq:#1})}

\newcommand{\be}{\begin{equation}}
\newcommand{\ee}{\end{equation}}
\newcommand{\bea}{\begin{eqnarray}}
\newcommand{\eea}{\end{eqnarray}}
\newcommand{\setl}{\setlength\arraycolsep{2pt}}

\newcommand{\noi}{\noindent}
\newcommand{\nn}{\nonumber}
\newcommand{\ra}{\rightarrow}
\newcommand{\Ra}{\Rightarrow}

\newcommand{\cH}{{\cal H}}

\newcommand{\cL}{{\cal L}}
\newcommand{\cM}{{\cal M}}

\newcommand{\cO}{{\cal O}}
\newcommand{\cP}{{\cal P}}

\newcommand{\cZ}{{\cal Z}}

\newcommand{\Imm}{\mbox{\rm Im}}
\newcommand{\Ree}{\mbox{\rm Re}}
\newcommand{\Tr}{\mbox{\rm Tr}}
\newcommand{\tr}{\mbox{\rm tr}}

\newcommand{\MeV}{\mbox{\rm MeV}}

\newcommand{\annd}{\mbox{\rm and}}
\newcommand{\foor}{\mbox{\rm for}}

\newcommand{\stern}{\langle\bar{\psi}\psi\rangle}

\newcommand{\Nc}{\mbox{${\rm N_c}$}}

\input epsf

\def\theequation{\arabic{section}.\arabic{equation}}

\begin{document}

\begin{titlepage}

\begin{flushright}
\end{flushright}

\vspace*{0.2cm}
\begin{center}
{\Large {\bf Large--$\Nc$ QCD and Harmonic Sums~\footnote{Based on my talk at  "Raymond's 80th Birthday Party", LAPP, July 11th  2011.} }}\\[2 cm]

{\bf Eduardo de Rafael}\\[1cm]

 {\it Centre  de Physique Th{\'e}orique~\footnote{Unit{\'e} Mixte de Recherche (UMR 6207) du CNRS et des Universit{\'e}s Aix Marseille 1, Aix Marseille 2 et Sud Toulon-Var, affili{\'e}e {\`a} la FRUMAM.}\\
       CNRS-Luminy, Case 907\\
    F-13288 Marseille Cedex 9, France}

\end{center}

\vspace*{3.0cm}

\begin{abstract}
In the Large--${\rm N_c}$ limit of  QCD,  two--point functions of local operators become Harmonic Sums. I review some properties which follow from this fact and which are relevant for phenomenological applications. This has led us to consider a class of Analytic Number Theory Functions as toy models of Large--${\rm N_c}$ QCD which I also discuss.
\end{abstract}

\end{titlepage}

\section{\normalsize Introduction}\lbl{int}
\setcounter{equation}{0}
\def\theequation{\arabic{section}.\arabic{equation}}

Many of us would like to know the answer to the following questions:

\begin{itemize}

	\item
	What is the effective Field Theory of QCD at Long Distances ?
	
	\item
	How does QCD fix the couplings of the Chiral Lagrangian of the Nambu-Goldstone modes of the spontaneously broken chiral--${\rm SU(3)}$ flavour symmetry~? 
	
 \item
 
Can we answer these questions, perhaps more easily,  within the framework of QCD in the Large--$\Nc$ limit~\cite{tH74}~? 

In that respect, it has been shown~\footnote{See refs.~\cite{VW74,tH79,CW80,VW84}. For a lucid exposition see ref.~\cite{Knecht95}.} that if the confinement property of QCD persists in this limit there is also spontaneous chiral symmetry breaking and the Hadronic Spectrum consists then of an infinite number of Narrow States~\cite{Wi79}.

\end{itemize}
\noi
Unfortunately, in spite of the successes of the Standard Model,  the answer to these questions remains unknown. What I shall do here is to provide a few comments related to them.

\subsection{\small General Comments}\lbl{int1}

\begin{enumerate}
	\item 
Independently of the Large--$\Nc$ approximation, the couplings of the effective Chiral Lagrangian of the Strong Interactions of the Nambu-Goldstone modes, can be identified with the  coefficients of  the Taylor expansion of appropriate QCD Green's Functions. 

\item
By contrast, most of the couplings of the effective chiral Lagrangian of the Electroweak  Interactions of the same Nambu-Goldstone modes are given by integrals over all the range of euclidean momenta of appropriate two--point functions with soft insertions of local operators. Their determination, therefore,  requires a precise matching of the short distance and the long distance contributions to the underlying QCD Green's functions. 
\end{enumerate}

\noi
Typical terms of the chiral Lagrangian are:

{\setl
\bea\lbl{chl}
\cL_{\footnotesize\rm eff} & = & \frac{1}{4}{ F_{\pi}^2} \underbrace{\tr\left( D_{\mu}U D^{\mu}U^{\dagger}\right)}_{\footnotesize \pi\pi\ra\pi\pi\,,~ K\ra \pi e \nu}+{\cL}_{\it WZW}+\underbrace{{ L_{10}}\ \tr\left( U^{\dagger}F_{R\mu\nu}UF_{L}^{\mu\nu}\right)}_{\footnotesize \pi\ra e\nu\gamma}+\cdots \nn \\
 & & \underbrace{ e^2 {C}\  \tr (Q_R U Q_L U^{\dagger})}_{\footnotesize -e^2 { C} \frac{2}{ F_{\pi}^2}(\pi^+\pi^- +\ K^{+}K^{-})}-\underbrace{\frac{G_{\footnotesize\rm F}}{\sqrt{2}}V_{ud} V_{us}^*\  g_{\underline 8}\  F_{\pi}^4 \left(D_{\mu}U D^{\mu}U^{\dagger}\right)_{23}}_{\footnotesize K\ra \pi\pi,~K\ra \pi\pi\pi} +  \cdots \,.
\eea}

\noi
Here $U$ denotes the $3\times 3$ unitary matrix in the $u,d,s$ flavour space which collects the Nambu--Goldstone fields and which under chiral rotations transforms as $U\ra V_R U V_L^{\dagger}$; $D_{\mu}U$ denotes the covariant derivative in the presence of external vector and axial-vector sources. 
The first term in the first line is the lowest order effective Lagrangian in the sector of the strong interactions~\cite{Wei79} with $F_{\pi}$ the pion--decay coupling constant in the chiral limit where the light quark masses $u,d, s$ are neglected ($F_{\pi}\simeq 90~\MeV$). The second term stands for the anomalous Wess--Zumino--Witten~\cite{WZ71,Wi83} effective Lagrangian of $\cO(p^4)$. The third one shows a typical term of $\cO(p^4)$ in the chiral counting~\cite{GL84}, with $L_{10}$ a coupling constant which is not fixed by symmetry requirements alone. The first term in the second line corresponds to the lowest order effective Lagrangian, which is $\left[\cO(p^0) \right]$ in the chiral counting~\cite{EGPdeR89}, and which appears when photons are integrated out in the presence of the strong interactions ($Q_L =Q_R ={\rm diag.[2/3,-1/3,-1/3]}$ and $e$ is the electric charge). The second term in the second line is the lowest order effective Lagrangian in the electroweak sector which induces non--leptonic $K$--decays~\footnote{$G_{\footnotesize\rm F}$ is the Fermi constant and $V_{ud}$, $V_{us}^*$  matrix elements of the flavour mixing matrix. The coupling $g_{\underline 8}$ governs the strength of the dominant $\Delta I=1/2$ transitions. For more  details see e.g. ref.~\cite{HPdeR03} and references therein.}. Typical physical processes to which each term contributes are indicated under the braces. Each term in the chiral Lagrangian is modulated by a coupling constant ($F_{\pi}^2\,, L_{10}\,,\cdots C\,, g_{\underline 8}\cdots $ in Eq.~\rf{chl}), which encodes the underlying dynamics responsible for the appearance of the corresponding effective term. The evaluation of these couplings from the underlying QCD Lagrangian is the question we are interested in.

\subsection{\small The Left--Right Correlation Function as an Example}\lbl{int2}

As a precise  example of the comments in the previous subsection let us consider   the left--right correlation function:
\be\lbl{LR}
\Pi_{\rm LR}^{\mu\nu}(q)=2i\int d^4 x\,e^{iq\cdot x}\langle 0\mid
T\left(L^{\mu}(x)R^{\nu}(0)^{\dagger}\right)\mid 0\rangle \,,
\ee with left and right currents:
\be L^{\mu}(x)=\bar{d}(x)\gamma^{\mu}\frac{1}{2}(1-\gamma_{5})u(x)
\qquad \annd \qquad
R^{\mu}(x)=\bar{d}(x)\gamma^{\mu}\frac{1}{2}(1+\gamma_{5})u(x)\,.
\ee
The discussion here, unless explicitly mentioned, does not use the Large--$\Nc$ approximation.
In the chiral limit where the light quark masses are set to zero, this
two--point function only depends on one invariant function ($Q^2=-q^{2}\ge 0$
for
$q^2$ spacelike)
\be
\Pi_{\rm LR}^{\mu\nu}(q)=(q^{\mu}q^{\nu}-g^{\mu\nu}q^2)\ \Pi_{\rm LR}(Q^2)\,.
\ee
The self--energy function
$\Pi_{\rm LR}(Q^2)$ in the chiral limit  vanishes order by order in QCD perturbation theory and is an
order parameter of the spontaneous breakdown of chiral symmetry for all
values of the momentum transfer~\cite{KdeR98}. In what follows we shall be working in this limit. 

The Taylor expansion of $\Pi_{\rm LR}(Q^2)$ at low $Q^2$ is a power series in $Q^2$:
\be
-Q^2\Pi_{\rm LR}(Q^2)=F_{\pi}^2+4L_{10}^{\rm\tiny eff}Q^2+\cO(Q^4)\,,
\ee
More precisely
\be
L_{10}^{\rm\tiny eff}  =  L_{10}(\mu)+{\rm Goldstone~one~loop~corrections}
\ee
where $L_{10}(\mu)$ denotes the $\cO(p^4)$ coupling in Eq.~\rf{chl}
renormalized at the scale $\mu$. In the $1/\Nc$--expansion, the Goldstone loop corrections are subleading and, therefore,  only the tree level couplings survive in the Large--$\Nc$ limit, which is one of the important simplifications of this limit.

The function $\Pi_{\rm LR}(Q^2)$ also governs the coupling constant $C$ in Eq.~\rf{chl} and gives a mass of  electromagnetic origin~~\cite{Lowetal67}  (and also from the integration of the heavy $Z$ electroweak boson~\cite{KPdeR98}) to the $\pi^{\pm}$ and $K^{\pm}$ particles:
\be\lbl{piem}
\frac{2e^2 C}{F_{\pi}^2}=
m_{\pi^{\pm}}^{2}=m_{K^{\pm}}^{2}=\frac{\alpha}{\pi}\,\frac{3}{8F_{\pi}^2}\,
\int_0^\infty
dQ^2\,\left[-Q^2\Pi_{\rm LR}(Q^2)\right]\ .
\ee
This integral converges in the ultraviolet
because~\cite{SVZ79}
\be\lbl{OPE}
\lim_{Q^2\ra\infty}\Pi_{\rm LR}(Q^2)\sim \cO\left(\frac{\stern^2}{Q^6}\right)\,.
\ee
Furthermore, it has also been shown~\cite{W83, CLT95} that
\be\lbl{witten}
-Q^2\Pi_{\rm LR}(Q^2)\ge 0 \qquad\foor~{\rm all}\  \quad 0\le Q^2\le\infty\,,
\ee
which in particular ensures the positivity of the integral in
Eq.~(\ref{eq:piem}) and thus the stability of the QCD vacuum with respect
to  small perturbations induced by electroweak interactions.

\section{\normalsize Large--$\Nc$ Models and Phenomenology}\lbl{models}
\setcounter{equation}{0}
\def\theequation{\arabic{section}.\arabic{equation}}

QCD models of spontaneously chiral symmetry breaking, like the Constituent Chiral Quark Model (C$\chi$QM) of Georgi and Manohar~\cite{MG84,EdeRT90,Wei10,EdeR11},  or the more sophisticated Extended Nambu--Jona-Lasinio model (ENJL)~\cite{NJL61,BBdeR93} have been rather successful in reproducing the phenomenological determinations of the couplings of the chiral Lagrangian in the strong interaction sector. However, they fail in general to provide the required matching between short and long distances which is needed in order to evaluate the couplings of the chiral Lagrangian induced by the Electroweak Interactions, i.e. couplings like $C$ and $g_{\underline 8}$ in Eq.~\rf{chl}. Because of this, in the phenomenological applications,  these models have progressively  been replaced by a more direct approach where the relevant Green's functions are approximated by a \underline{finite} number of the Large--$\Nc$ QCD hadronic spectrum of narrow states. In fact, the
Resonance Chiral Lagrangians  of the type discussed in refs.~\cite{EGPdeR89,EGLPdeR89} and their extensions~(see e.g. ref.~\cite{CENP03} and references therein), can be viewed as  simplified versions of the Large--${\rm N_c}$ QCD Hadronic Lagrangian when limited to a \underline{finite} number of states.
 
The methodology which has been suggested~(see e.g. ref.~\cite{EdeR03} for a review), consists in fixing the couplings and masses of a Minimal Hadronic Ansatz (MHA) of narrow states which contribute to a specific Green's function in such a way that, on the one hand the short distance behaviour predicted by the operator product expansion (OPE)~\cite{SVZ79} of the underlying Green's function in Large--${\rm N_c}$ QCD is satisfied and, on the other hand,  the long distance behaviour constraints governed by the effective chiral Lagrangian in the sector of the Strong Interactions alone are satisfied as well. 
As an example of this MHA approach, let us consider again the integral in Eq.~ \rf{piem} which fixes the coupling $C$. The MHA in this case requires the presence of three states: the massless pion pole, a vector state with mass $M_V$ and  an axial-vector state with mass $M_A$. The constraint that $\Pi_{\rm LR}(Q^2)$ satisfies the OPE at short distances (see Eq.~\rf{OPE}) implies that $\cO(1/Q^2)$ terms and $\cO(1/Q^4)$ terms must be absent. This, as we shall later discuss, also implies the validity of so called Weinberg sum rules, and fixes the couplings of the $V$ and $A$ states in the spectral function of the left--right correlation function with the result
\be\lbl{MHA}
-Q^2\Pi_{LR}(Q^2)=F_{\pi}^{2}\frac{M_{V}^2
M_{A}^2}{(Q^2+M_{V}^2)(Q^2+M_{A}^2)}\,.
\ee
Inserting this function in Eq.~\rf{piem} gives a prediction for the
$\pi^{+}-\pi^{0}\equiv\Delta m_{\pi}$ mass difference~\footnote{This is the result for
$F_{\pi}\!=\!(87\pm 3.5)\,\MeV$, $M_{V}\!=\!(748\pm 29)\,\MeV$ and
$g_{A}\!=\!\frac{M_{V}^2}{M_{A}^2}\!=\!0.50\pm 0.06$. These values follow
from an overall fit to predictions of the low energy
constants in the MHA to Large--$\Nc$ QCD.}:
\be\lbl{mpimha}
\Delta m_{\pi}=(4.9\pm 0.4)\,\MeV\,, 
\ee
to be compared with the experimental value:
$
\Delta m_{\pi}=(4.5936\pm 0.0005)\,\MeV 
$.

The MHA approach to Large--$\Nc$ QCD has led to a remarkable set of interesting predictions for some of the couplings of the Electroweak Lagrangian in the chiral limit~(two representative references are~\cite{PdeR00,HPdeR03}). The incorporation of chiral corrections, however, becomes technically rather cumbersome and, above all, the question of the reliability of the approximation with a {\it finite} number of narrow states to Large--${\rm N_c}$ QCD, to which I shall later come back, remains open (see e.g. ref.~\cite{MP07}).

\subsection{\small Comment on Minkowsky versus Euclidean}\lbl{models1}

The spectral function of the left--right correlation function defined in
Eq.~\rf{LR} can be obtained from measurements of the hadronic
$\tau$--decay spectrum (vector--like decays minus axial--vector like
decays). Figure~1 below shows the experimental
determination of
$\frac{1}{\pi}\Imm\Pi_{\rm LR}(t)$, obtained from the ALEPH
collaboration~\cite{ALEPH} data at LEP, versus the invariant hadronic mass
squared $t$ in the accessible region $0\le t\le m_{\tau}^2$. This plot is a good experimental proof of spontaneously chiral symmetry breaking in Nature. If the symmetry was realized \`a la Wigner--Weyl, the shape should be a straight horizontal line all the way down to zero. 
\begin{figure}[h]

\begin{center}
\includegraphics[scale=0.8]{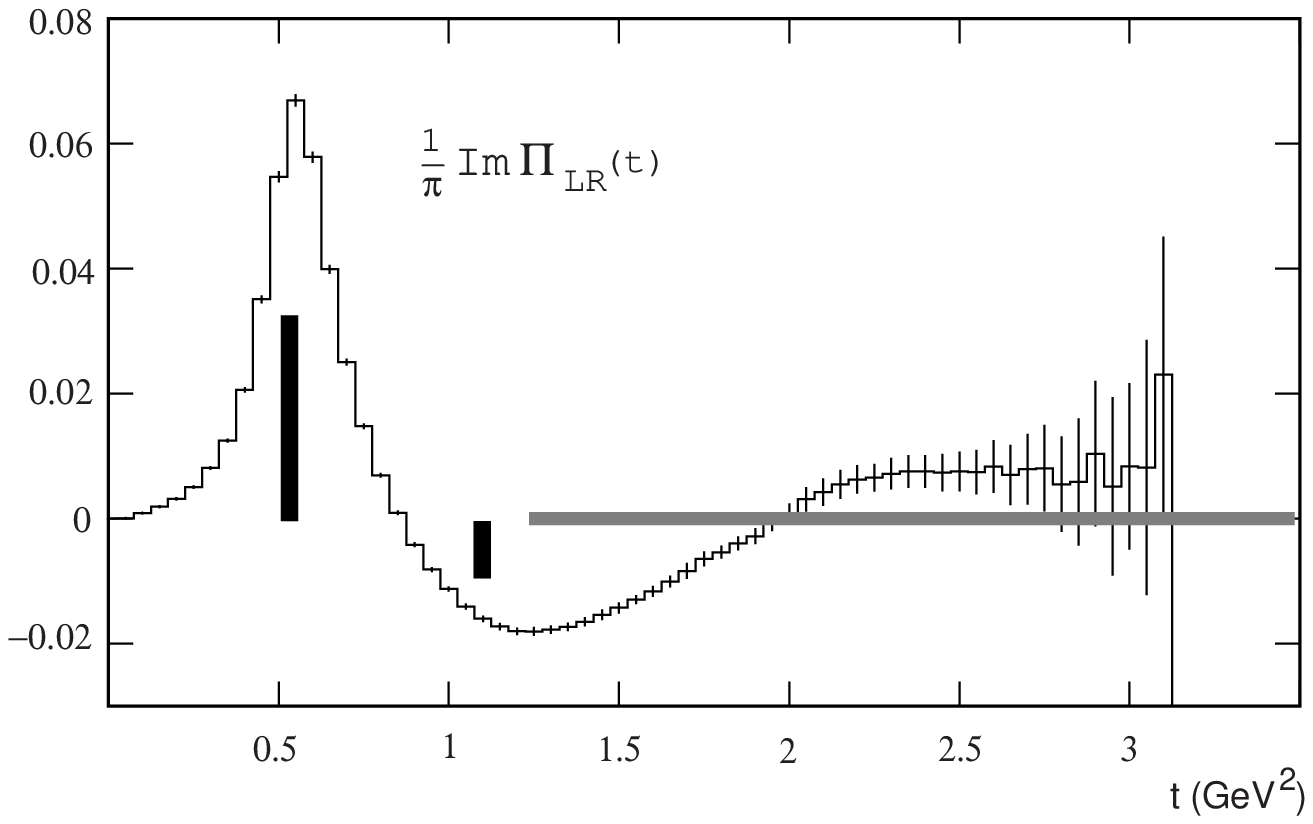}
\end{center}

\vspace*{0.25cm}
{\bf Fig.~1}
{\it\small The Spectral Function
$\frac{1}{\pi}\Imm\Pi_{\rm LR}(t)$ compared to the MHA to Large--$\Nc$ QCD. }
\lbl{fig:ExTh}

\end{figure}
\noi
Also shown in the same Fig.~1 is the simple spectrum of the MHA approximation to Large--$\Nc$
QCD which, as discussed before, consists of the pion pole (not shown in the
figure), a vector narrow state (the first positive vertical line)  and an axial--vector
narrow state the second negative vertical line). The heights of these vertical lines represent the strengths of the couplings of the narrow states in the spectral function. At this level of approximation, and in the
chiral limit, the rest of the vector and axial--vector states are degenerate and, therefore,  
they  cancel in the difference, a fact which in any case is predicted by QCD at high energies when perturbation theory applies, and is
reproduced by the asymptotic horizontal continuum line shown in the figure.  

Looking
at the plot in Fig.~1, one can hardly claim that the MHA approximation in the Minkowsky region reproduces
the details of the experimental data. This is to be contrasted, however, with what happens in the Euclidean region. With 
$\Pi_{\rm LR}(Q^2)$ determined from the spectral function by the unsubtracted
dispersion relation
\begin{figure}[h]

\begin{center}
\includegraphics[scale=0.5]{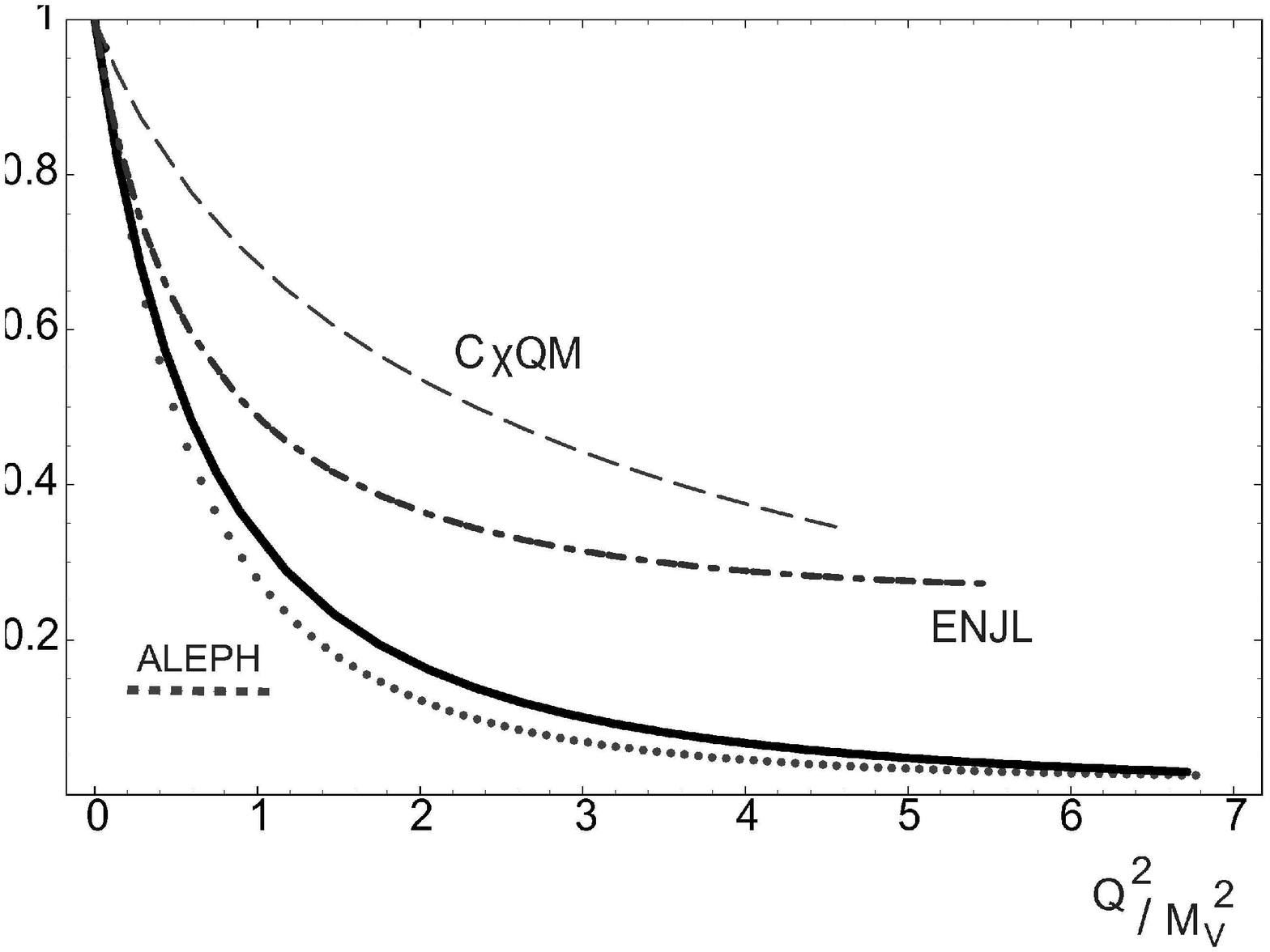}
\end{center}

{\bf Fig.~2}
{\it\small Plot of $\frac{-Q^2}{F_{\pi}^2}
\Pi_{\rm LR}(Q^2)$ in the Euclidean region. The
solid curve is the one corresponding to the MHA to Large--$\Nc$ QCD  and the dotted curve the one from the experimental data
in Fig~1. The other curves are the predictions of the Constituent chiral Quark Model(C$\chi$QM)  and the Extended Nambu--Jona-Lasinio Model (ENJL).}

\lbl{fig:EuLRAA}

\end{figure}

\be\lbl{displr}
\Pi_{\rm LR}(Q^2)=\int_{0}^{\infty}dt\frac{1}{t+Q^2}
\frac{1}{\pi}\Imm\Pi_{\rm LR}(t)\,,
\ee 
the corresponding plot of the function
$\frac{-Q^2}{F_{\pi}^2}
\Pi_{\rm LR}(Q^2)$ versus the Euclidean variable $Q^2$ normalized to vector mass $M_V ^2$, as shown in
Fig.~2 (the solid curve), reproduces rather well the dotted curve, which is the one resulting from the experimental
data in Fig.~1. This is in fact a generic feature: Green's functions
in the Minkowski region have  a lot of structure, while the corresponding shape  in the Euclidean region shows a smooth behaviour. As shown in this example, the simple
MHA in the euclidean region already provides a rather good interpolation between the asymptotic
regimes where, by construction, it has been constrained to satisfy the
lowest order chiral behaviour and the leading and next to leading OPE constraints. This  good interpolation is the reason why the integral in Eq.~\rf{piem}, evaluated in the MHA, already reproduces the experimental
result rather well. One can also see in Fig.~1 that models which fail to incorporate the matching to short distances, like the C$\chi$QM and the ENJL model, fail to reproduce the shape of the experimental curve already at rather low values of $Q^2$.

\subsection{\small More on the C$\chi$QM}\lbl{models2}

Concerning the failure of the matching between long and short distances in Models of Large--$\Nc$ QCD,  I would like to point out that 
there is, however,  a class of  low--energy observables, governed by integrals of specific QCD Green's functions, for which  the C$\chi$QM predictions encoded in the  MG--Lagrangian, in spite of its limitations, can be rather reliable. This is the case when the 
leading short distance behaviour of the underlying Green's functions of a given observable is governed by perturbative QCD. Interesting examples of this class of observables   are the decay rate $\pi^0 \ra e^+ e^-$, the Hadronic Vacuum Polarization and the Hadronic Light--by--Light Scattering contributions to a low--energy observable like the anomalous magnetic moment of the muon: $\frac{1}{2}(g_{\mu}-2)$. Furthermore, as recently pointed out by Weinberg~\cite{Wei10}, the MG--Lagrangian in the Large--${\rm N_c}$ limit,  modulo the addition of a finite number of local counterterms~\cite{EdeR11}, is a renormalizable Lagrangian. Calculations with the MG--Lagrangian, compared to those with the more sophisticated approaches described above, have the advantage of simplicity and, when applied to this class of low--energy observables, can provide a check to the more elaborated phenomenological approaches. The fact that one is dealing with a renormalizable Quantum Field Theory is, of course, a welcome feature.

The C$\chi$QM effective Lagrangian in question is the following:

{\setl
\bea\lbl{CCQL}
\cL_{{\rm C}\chi{\rm QM}}(x) & \!\! =\!\! & \underbrace{i{\bar Q}\gamma^{\mu}\left(\partial_{\mu}+\Gamma_{\mu}+iG_{\mu} \right)Q-\frac{i}{2}{g_A}\ {\bar Q}\gamma^{\mu}\gamma_5 \xi_{\mu}Q-M_{Q} {\bar Q}Q}_{\it M-G}-\frac{1}{2}{\bar Q}\left(\Sigma -\gamma_5 \Delta \ \right)Q \nn\\ 
 & + & \underbrace{\frac{1}{4}{ F_{\pi}}^2 \tr\left[ D_{\mu}UD^{\mu}U^{\dagger}\right.}_{\it M-G}
+\left. U^{\dagger}\chi+\chi^{\dagger}U\right]-\underbrace{\frac{1}{4}\sum_{a=1}^{8} G_{\mu\nu}^{(a)}G^{(a)\mu\nu}}_{\it M-G}+ e^2 { C}\ \tr (Q_R U Q_L U^{\dagger}) \nn \\
 & + & 
{ L_5}\  \tr D_{\mu}U^{\dagger}D^{\mu}U(\chi^{\dagger}U+U^{\dagger}\chi) +
{ L_8}\  \tr (U\chi{\dagger}U\chi{\dagger}+U^{\dagger}\chi U^{\dagger}\chi)\,.
\eea}

\noi
The underbraced terms are  those of the MG--Lagrangian, but in the presence
of external $SU(3)$ vector $v_{\mu}(x)$ and axial-vector $a_{\mu}(x)$ sources. The field matrix $U(x)$ is the same  3$\times$3 unitary matrix in flavour space which collects the Goldstone fields as in Eq.~\rf{chl}. The vector field matrix $D_{\mu}U$ is also the same covariant derivative of $U$ as in Eq.~\rf{chl} with respect to the same external sources:
\be
D_{\mu}U=\partial_{\mu}U-ir_{\mu}U+iUl_{\mu}\,,\quad l_{\mu}=v_{\mu}-a_{\mu}\,,\quad r_{\mu}=v_{\mu}+a_{\mu}\,, 
\ee 
and, with $U=\xi\xi$,
\be
\Gamma_{\mu}=\frac{1}{2}\left[
\xi^{\dagger}(\partial_{\mu}-ir_{\mu})\xi+
\xi(\partial_{\mu}-il_{\mu})\xi^{\dagger}\right]\,,\quad
\xi_{\mu}=i\left[\xi^{\dagger}(\partial_{\mu}-ir_{\mu})\xi-
\xi(\partial_{\mu}-il_{\mu})\xi^{\dagger}\right]\,.
\ee
The gluon field matrix in the fundamental representation of color $SU(3)$ is $G_{\mu}(x)$ and $G_{\mu\nu}^{(a)}(x)$ its corresponding gluon field strength tensor.
The presence of external scalar $s(x)$ and pseudoscalar $p(x)$ sources  induces the extra terms proportional to
\be
\chi= 2 B [s(x)+i p(x)]\,,
\ee
where $B$, like $F_{\pi}$, is an order parameter which has to be fixed from experiment. When these sources are frozen to the up, down, and strange light quark masses of the QCD Lagrangian, 
\be
\chi= 2 B \cM\,,\quad {\rm with}\quad \cM={\rm diag}(m_u\,, m_d\,, m_s)\,,
\ee
and then
\be
\Sigma=\xi^{\dagger}\cM \xi^{\dagger}+\xi\cM^{\dagger} \xi\,,\quad 
\Delta=\xi^{\dagger}\cM \xi^{\dagger}-\xi\cM^{\dagger}\xi\,.
\ee

With the axial coupling  fixed to  $g_A =1$,
the extra couplings $L_5$ and $L_8$
are the only terms which are needed to absorb the ultraviolet (UV) divergences  when the constituent quark fields $Q(x)$ are integrated out~\footnote{We disregard divergent couplings involving external fields alone to lowest order in the chiral expansion.}. If one wants to consider the case where photons are also integrated out then, to leading order in the chiral expansion and in the electric charge coupling $e$, the last term in the second line is also required to absorb further UV--divergences. 
Loops involving pion fields are subleading in the $1/{\rm N_c}$--expansion and hence, following the observation of Weinberg in ref.~\cite{Wei10}, the Lagrangian in Eq.~\rf{CCQL}, when considered within the framework of the large--${\rm N_c}$ limit, is a renormalizable Lagrangian.

\section{\normalsize General Properties of Large--$\Nc$ QCD}\lbl{lncqcd}
\setcounter{equation}{0}
\def\theequation{\arabic{section}.\arabic{equation}}

An interesting feature of Large--$\Nc$  QCD which I shall discuss next is the fact that  two--point functions of color singlet  composite local operators become simple Harmonic Sums in this limit.

 In full generality~\footnote{For a clear introduction to the appropriate mathematics literature see e.g. ref.~\cite{FGD95}.}, an Harmonic Sum is characterized by a {\it Base Function} (in our case the kernel in the dispersion relation which the two--point function in question obeys) and a {\it Dirichlet series}.
\begin{equation}\label{dirich}
	\Sigma(s)=\sum_{n=1}^{\infty}\lambda_n \mu_n^{-s}\,,
\end{equation}
where the $\mu_n$ are called the {\it Frequencies} (the position of the narrow states in the Minkowski region in our case) and the $\lambda_n$ the {\it Amplitudes} (the residues of the corresponding poles). 

The simplest example of a {\it Dirichlet Series} is the {\it Riemann zeta function}:
\begin{equation}\label{Rie}
	\zeta(s)=\sum_{n=1}^{\infty} n^{-s}= \prod_{{\rm primes}(p)}\frac{1}{1-\frac{1}{p^s}}\,,\quad {\rm Re}(s)>1\,,
\end{equation}
where $\lambda_n =1$ and $\mu_n =n$.  The Euler product expression in the r.h.s.  extends to all prime numbers $p$.

\subsection{\small The $\Pi_{\rm LR}(Q^2)$ Self--Energy as an Harmonic Sum}
\noi
Let us consider again  the left--right correlation function in the chiral limit as an example to illustrate these properties.
In the Large--$\Nc$ limit, the spectral function $\frac{1}{\pi}\Imm\Pi_{\rm LR}(t)$ is the sum of an infinite number of narrow states
\begin{equation}\label{spectral}
\frac{1}{\pi}\Imm\Pi_{\rm LR}(t)=-F_{\pi}^2\delta(t)+\sum_{n=1}^{\infty}(-1)^{n+1} F_n^2\  \delta(t-M_n^2)\,.
\end{equation}
with positive weights for the vector--like components and negative weights for the axial-vector--like components. The first term is the contribution from the pion pole.
Inserting this spectral function in the dispersion relation in Eq.~\rf{displr} we find
\begin{equation}
	-\frac{Q^2}{F_{\pi}^2}	\Pi_{\rm LR} (Q^2)  
		 = 1- \sum_{n=1}^{\infty}\  (-1)^{n+1} \frac{F_{n}^2}{F_{\pi}^2} \frac{1}{1+\frac{M_n^2}{Q^2}}\,.
\end{equation}
This is a typical Harmonic Sum which we can write as follows: 
\begin{equation}
	1+\frac{Q^2}{F_{\pi}^2}	\Pi_{\rm LR} (Q^2)  
		 = \sum_{n=1}^{\infty} \lambda_n \ B_{\rm LR}\left(\mu_n \frac{M_{\rho}^2}{Q^2} \right)\,,
\end{equation}
where ($M_{\rho} \equiv M_1$)
\begin{equation}\label{lambdamu}
\lambda_n = (-1)^{n+1} \frac{F_n^2}{F_{\pi}^2}\quad\annd\quad \mu_n =\frac{M_n^2}{M_{\rho}^2}
\end{equation}
and where the {\it Base Function} is
\begin{equation}
	B_{\rm LR}(x)=\frac{1}{1+x}\,.
\end{equation}
 
The crucial property of Harmonic Sums is that they have a factorizable Mellin--Transform~\footnote{The Mellin transform of a function $F(z)$: $\cM[F(z)](s)$, is defined by the integral $\cM[F(z)](s)=\int_0^\infty dz z^{s-1}F(z)$, in the domain of $s$ where the integral exists and elsewhere by its analytic continuation.}.
In our case
\begin{equation}\lbl{mtlr}
\cM\left[1+\frac{Q^2}{F_{\pi}^2}	\Pi_{\rm LR} (Q^2)  \right](s)= \cM[B_{\rm LR}(x)](s)\ \Sigma(s)\,,	
\end{equation}
with
\begin{equation}\label{MBR}
\cM[B_{\rm LR}(x)](s)=\Gamma(s)\Gamma(1-s)\,,
\end{equation}
and $\Sigma(s)$  the Dirichlet series in Eq.~\ref{dirich} with  $\lambda_n$ and $\mu_n$ defined in Eqs.~\ref{lambdamu}. Inverting the Mellin transform in Eq.~\rf{mtlr} results then in the following Mellin--Barnes representation:
\begin{equation}\lbl{AMB}
	-\frac{Q^2}{F_{\pi}^2}	\Pi_{\rm LR} (Q^2) =1-\frac{1}{2\pi i}\int\limits_{c-i\infty}^{c+i\infty} ds\  \left(\frac{M_{\rho}^2}{Q^2}\right)^{-s}\  \Sigma(s)\ \Gamma(s)\Gamma(1-s)\,, 
\end{equation}
where the integration path lies in the so called  {\it Fundamental Strip}~\cite{FGD95} which is defined by the intersection of the convergence domain of the  base function  ($\rm {Re}(s)\;\in\;]0,+1[$ in our case), with the domain of absolute convergence of the Dirichlet Series $\Sigma(s)$. 
Notice that in full generality, independently of the Large--$\Nc$ approximation, $\Pi_{\rm LR} (Q^2)$  also has a Mellin--Barnes representation like the one in Eq.~\rf{AMB}, except that then $\Sigma(s)$ is the Mellin transform of the spectral function $\frac{1}{\pi}\Imm\tilde{\Pi}_{\rm LR}(t)$ where the {\it tilde} means that the contribution from the pion pole has been removed:
\be\lbl{mellincont}
\Sigma(s)=\frac{M_{\rho}^2}{F_{\pi}^2}\int_0^\infty \frac{dt}{M_{\rho}^2}\  \left(\frac{t}{M_{\rho}^2} \right)^{-s}\frac{1}{\pi}\Imm\tilde{\Pi}_{\rm LR}(t)\,.
\ee
This integral, in the Large--$\Nc$ limit, becomes the Dirichlet Series:
\be
\Sigma(s)=\int_0^\infty \frac{dt}{F_{\pi}^2}\left(\frac{t}{M_{\rho}^2} \right)^{-s}\frac{1}{\pi}\Imm\tilde{\Pi}_{\rm LR}(t)\underset{{{\rm Large-N_c}}}{\Ra}\ 
\sum_{n=1}^{\infty}(-1)^{n+1}\frac{F_{n}^2}{F_{\pi}^2}\left(\frac{M_{n}^2}{M_{\rho}^2}\right)^{-s}\,.
\ee
  
The  Mellin--Barnes representation is very useful. All the dynamics is encoded in the factorized Dirichlet series $\Sigma(s)$ or, in full generality, in the Mellin transform of the Spectral Function. The dependence in $Q^2$  is completely factorized from the details of the spectrum, and one can read off the asymptotic behaviours for small $Q^2$ (i.e. the  chiral expansion) and large--$Q^2$ (i.e. the short distance expansion or  OPE expansion) in a straightforward way~\footnote{Applications of this property of the Mellin--Barnes representation in QED and QCD have been recently discussed in refs.~\cite{FGdeR06,AGdeR08,GP10} and references therein.}. These asymptotic behaviours can be obtained from the so called {\it inverse mapping theorem}~\cite{FGD95} as follows:

\begin{itemize} 

\item
The coefficients $\mathsf{a}_{\mathsf{p},k}$ of the {\it singular series expansions} at the {\it left} of the {\it fundamental strip} i.e.
\be
{\Sigma(s)}\ \Gamma(s)\Gamma(1-s)\asymp\sum_{\Re\mathsf{p}\geq 0}\ \sum_{k=0}^{N(\mathsf{p})}\ \frac{	\mathsf{ a}_{\mathsf{p},k}}{(s+\mathsf{p})^{k+1}}
\ee
govern the {\it asymptotic expansion}:
\be\lbl{asope}
-\frac{Q^2}{F_{\pi}^2}\Pi_{\rm LR}(Q^2)\underset{{Q^2\ \ra\  \infty}}{\thicksim} 1- 
\sum_{\Re\mathsf{p}\geq 0}\ \sum_{k=0}^{N(\mathsf{p})}\  \frac{(-1)^{k}}{k!}\ \mathsf{a}_{\mathsf{p},k}\ \left(\frac{M_{\rho}^2}{Q^2}\right)^{\mathsf{p}}\ \log^{k}\frac{M_{\rho}^2}{Q^2}\,.
\ee
\item
The coefficients
 $\mathsf{b}_{\mathsf{p},k}$ of the {\it singular series expansions} at the {\it right} of the {\it fundamental strip} i.e.
\be
{ \Sigma(s)}\ \Gamma(s)\Gamma(1-s) \asymp\sum_{\Re\mathsf{p}\geq1}\ \sum_{k=0}^{N(\mathsf{p})}\frac{	\mathsf{b}_{\mathsf{b}_{\mathsf{p},k}}}{(s-\mathsf{p})^{k+1}}
\ee
govern the {\it asymptotic expansion}:
\be\lbl{chiralas}
\quad -\frac{Q^2}{F_{\pi}^2}
{ \Pi}_{\rm LR}(Q^2)\underset{{{Q^2}\ \ra\  0}}{\thicksim}\ 1-\frac{(-1)^{k+1}}{k!} \sum_{\Re\mathsf{p}\geq 1}\sum_{k=0}^{N(\mathsf{p})}\ \mathsf{b}_{\mathsf{p},k}\ \left(\frac{M_{\rho}^2}{Q^2}\right)^{-\mathsf{p}}\log^k \frac{M_{\rho}^2}{Q^2}\,.
\ee
\end{itemize}

\noi
The sum structure over the poles $\mathsf{p}$ in the r.h.s. of these equations is completely general for any two--point function. When applied to the left--right correlation function there are, however,  further  specific restrictions to take into account:

\begin{itemize}

\item
The fact that in the OPE of ${ \Pi}_{\rm LR}(Q^2)$ there are no operators of dimension two and four implies that   
\be
\mathsf{ a}_{\mathsf{0},0}=1\,,\quad\annd\quad\mathsf{ a}_{\mathsf{1},0}=0\,,
\ee
and hence 
\be
\Sigma(0)=1\,,\quad\annd\quad \Sigma(-1)=0\,.
\ee
This is precisely the content of the Weinberg Sum Rules, conventionally written as follows:
\be
\int_0^\infty \frac{dt}{F_{\pi}^2}\frac{1}{\pi}\Imm\tilde{\Pi}_{\rm LR}(t)=1\,, \quad {\rm 1st~Weinberg~Sum~Rule}\,,
\ee
and
\be
\int_0^\infty \frac{dt}{F_{\pi}^2}\frac{t}{M_{\rho}^2}\frac{1}{\pi}\Imm\tilde{\Pi}_{\rm LR}(t)=0\,, \quad {\rm 2nd~Weinberg~Sum~Rule}\,.
\ee

Notice, however, that in full generality  sum rules like
\be
\Sigma(s)=\int_0^\infty \frac{dt}{F_{\pi}^2}\left(\frac{t}{M_{\rho}^2}\right)^{-s}\ \frac{1}{\pi}\Imm\tilde{\Pi}_{\rm LR}(t)=0\,,
\ee
at the {\it left of the fundamental strip} where $s\leq 0$, have to be understood as analytic continuations of the Mellin transform of the spectral function (or the  Dirichlet series in the Large--$\Nc$ approximation) in the region of $s$ of absolute convergence. It is in this sense that the two Weinberg sum rules should be understood.     
\item
Because of the absolute convergence of $\Sigma(s)$ at the {\it right of the fundamental strip}, all the poles in the corresponding series expansion must be simple poles, hence  $N(\mathsf{p})=0$ in Eq.~\rf{chiralas}, which means that the expansion at small $Q^2$ is a pure power series.

\end{itemize}

\section{\normalsize Toy Models of Large--$\Nc$ QCD}\lbl{lncqcd}
\setcounter{equation}{0}
\def\theequation{\arabic{section}.\arabic{equation}}

As we have seen, the phenomenological applications of Large--$\Nc$ QCD we have discussed in Section~II rely on the limitation of the hadronic spectrum to a {\it finite}  number of narrow states. In order to test this approximation and in the absence of a solution of QCD in the Large--$\Nc$ limit, one has to resort to models to investigate the issue~\footnote{Early work on models of this type can be found in refs.~\cite{BSZ98,GPPdeR02,CGP05} and references therein.}. As we shall see, some of these models bring in functions of Analytic Number Theory with quite interesting features. 

\subsection{\small The Hurwitz Model $\Pi_{\rm LR}(Q^2)$ }
\noi
This is a model where the spectral function of the left--right correlation function consists of an infinite number of narrow states 
\begin{equation}
	\frac{1}{\pi}\Imm\Pi_{\rm LR}(t)=-F_{\pi}^2 \delta(t) +F_V^2\sum_{n=0}^\infty \delta(t-M_V^2 -n\sigma^2)-F_A^2\sum_{n=0}^\infty \delta(t-M_A^2 -n\sigma^2)\,, 
\end{equation}
with the  vector states and the axial-vector states  equally spaced by an amount $\sigma$.
Implementing the two Weinberg sum rules in this spectral function reduces the number of independent parameters to
$F_{\pi}$, the lowest vector mass $M_V$ and the value of $g_A=\frac{M_V^2}{M_A^2}$, with $M_A$ the lowest axial-vector states. The other parameters are then fixed as follows:
\be
F_V^2 =F_A^2=F_{\pi}^2\ \frac{1+g_A}{1-g_A}\,,\quad \sigma^2=M_V^2\left(1+\frac{1}{g_A}\right)\,,\quad  \annd\quad g_A=\frac{M_V^2}{M_A^2} \,.
\ee
Notice that this forces the inequality
\be
g_A<1\quad{\rm and~therefore}\quad M_A^2 > M_{V}^2\,.
\ee
The Mellin--Barnes Representation in this model is then:

{\setl
\bea\lbl{LRMB}
	\lefteqn{-\frac{Q^2}{F_0^2}\Pi_{\rm LR}(Q^2)  =  1-
	\frac{g_A}{1-g_A}\ \frac{1}{2\pi i}\int\limits_{c-i\infty}^{c+i\infty}ds\ \left(1+\frac{1}{g_A}\right)^{1-s}\  \times}\nn \\
	 & &  
\left\{\zeta\left(s, \frac{g_A}{1+g_A}\right)-\zeta\left(s,\frac{1}{1+g_A } \right)\right\}\left(\frac{M_V^2}{Q^2}\right)^{-s}\ \Gamma(s)\Gamma(1-s)\,,
\eea}

\noi
where $\zeta(s,v)$ is the Hurwitz function, a generalization of the Riemann zeta function, defined by the series:
\be
\zeta(s,v)=\sum_{n=0}^{\infty}\frac{1}{(n+v)^s}\,,\quad {\rm Re} s>1\,, \quad {\rm with}\quad v~{\rm a~fixed~real~number}\quad 0< v\le 1\,,
\ee
and its analytic continuation. 
For $v=1$ it reduces to the Riemann zeta function.

The Hurwitz function is a special case of the so called Dirichlet L-functions~\cite{Ap76}. It has an integral representation in terms of a Mellin transform:
\be
\Gamma(s)\zeta(s,v)=\int_0^\infty dx\ x^{s-1}\frac{e^{-vx}}{1-e^{-x}}\,,\quad {\rm Re} s>0\,,
\ee
which sets the basis for its analytic continuation~\cite{Ap76}. The so defined $\zeta(s,v)$ is analytic for all $s$ except for a simple pole at $s=1$ with residue 1. 

\begin{figure}[h]

\begin{center}
\includegraphics[width=0.8\textwidth]{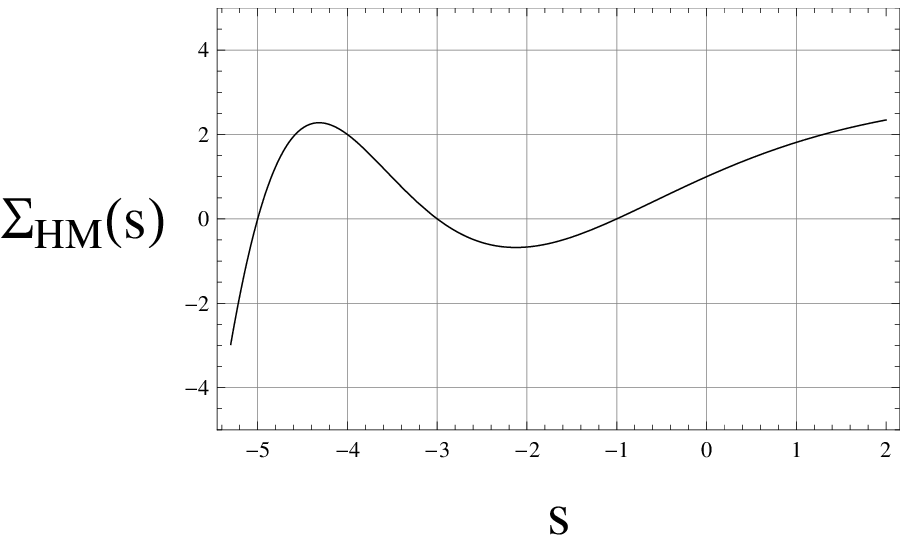}

\vspace*{0.25cm}
{\bf Fig.~3}
{\it\small  Plot of the Mellin Transform of the Hurwitz--Model in Eq.~\rf{mellinhm} for $g_A =1/2$.
}
\end{center}
\end{figure}

\noi
Comparing Eq.~\rf{LRMB} with Eq.~\rf{AMB}, we see that the relevant Mellin transform of the Hurwitz--Model is
\be\lbl{mellinhm}
\Sigma_{\rm HM}(s)=\frac{g_A}{1-g_A}\left(1+\frac{1}{g_A}\right)^{1-s}
\left[\zeta\left(s, \frac{g_A}{1+g_A}\right)-\zeta\left(s,\frac{1}{1+g_A } \right)\right]\,.
\ee
A plot of this Mellin transform is shown in Figs.~3.
Notice the zeros at $s=-1, -3, \cdots$. These zeros are in fact quite generic of the model and they occur at odd negative values of $s$. The origin of it is due to the property that, for $s=-m$ with $m=0,1,2,\dots$:
\be
\zeta(-m,v)=-\frac{B_{m+1}(v)}{m+1}\,,
\ee
where $B_{m+1}(v)$ denotes the Bernoulli polynomial of degree $m+1$. This fact plus the symmetry property
\be
B_{m}(1-x)=(-1)^m B_{m}(x)\,,
\ee
is at the origin of the zeros of the Mellin transform at $s=-1,-3,-5,\dots$. Notice also that the values of the Mellin transform at negative even integer values $s=-2,-4,-6,\dots$ are fixed by the values of Bernoulli polynomials $(m=0,1,2,3,\dots)$:
\be
\Sigma_{\rm HM}(-2m)=\frac{F_0^2}{M_V^2}\frac{g_A}{1-g_A}\left(1+\frac{1}{g_A} \right)^{1+m}\frac{1+(-1)^m}{m+1}B_{m+1}\left(\frac{1}{1+g_A}\right)\,.
\ee

\begin{figure}[h]

\begin{center}
\includegraphics[width=0.7\textwidth]{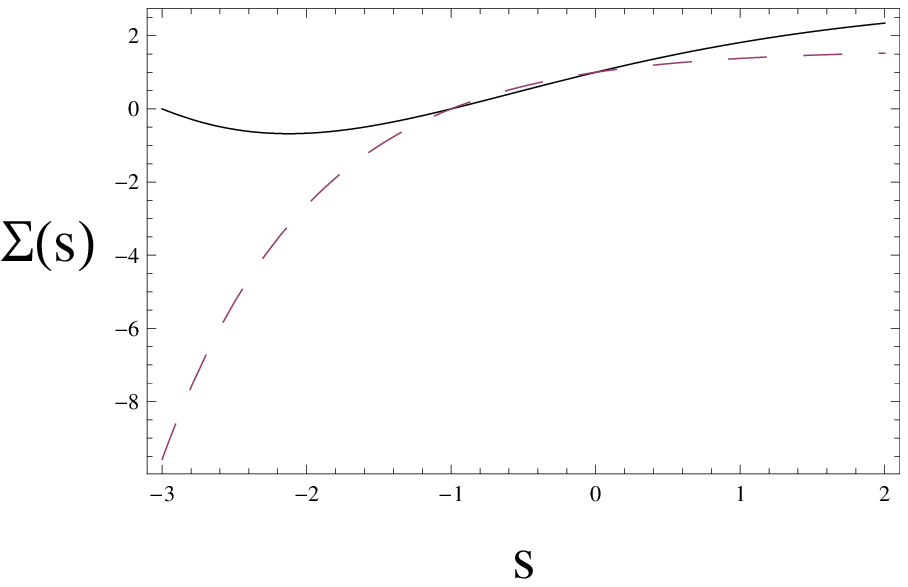}

\vspace*{0.25cm}
{\bf Fig.~4}
{\it\small  Mellin Transforms of the Hurwitz Model (solid curve) and MHA (dashed curve).
}
\end{center}
\end{figure}

At this stage, 
it is interesting to compare the Mellin transform of the Hurwitz model with the one of the MHA--approximation discussed in Section~II. This is shown in Fig.~4.
What emerges from this comparison is the fact that the two Mellin transforms do not differ much in the region $-1.5\lesssim s\lesssim 1.5$, the reason being that they both satisfy the two Weinberg sum rules. The result that the pion mass difference was rather well reproduced with the MHA approximation in Eq.~\rf{mpimha} is due to the fact that this observable is governed  by the slope of the Mellin transform at $s=-1$:
\be
\frac{d}{ds}\Sigma(s)\big\vert_{s=-1}=-\int_0^\infty \frac{dt}{F_{\pi}^2}\frac{t}{M_V^2} \log\left(\frac{t}{M_V^2}\right)\frac{1}{\pi}\Imm\tilde{\Pi}_{\rm LR}(t)\,.
\ee
The two Mellin transforms in Fig.~4 have, however, very different behaviors outside the interval $-1.5\lesssim s\lesssim 1.5$; not so much for positive $s$--values (corresponding to the chiral expansion), but they differ quite dramatically for negative $s$--values (corresponding to the OPE expansion) where  one finds: $\left(z=\frac{Q^2}{M_V^2}\right)$:
\be\lbl{condmha}
-\frac{Q^2}{F_{\pi}^2}\Pi_{\rm LR}(Q^2)\big\vert_{\rm MHA}  \underset{{z\rightarrow\  \infty}}{\thicksim}   \frac{2}{z^2}-\frac{6}{z^3}+\frac{14}{z^4}-\frac{30}{z^5}+\frac{62}{z^6}+
\cO\left(\frac{1}{z^7} \right)\,,
\ee
while
\be\lbl{condhm}
-\frac{Q^2}{F_{\pi}^2}\Pi_{\rm LR}(Q^2)\big\vert_{\rm HM}  \underset{{z\rightarrow\  \infty}}{\thicksim}  \frac{2}{3z^2}-\frac{2}{z^4}+\frac{14}{z^6}
+\cO\left(\frac{1}{z^{8}} \right)\,.
\ee
We can see  that, except for the first term, the signs are different and the discrepancy in magnitude increases for each successive term. Yet, the shapes of the two correlation functions in the Euclidean, as seen in Fig.~5, is not so different.   

\begin{figure}[h]

\begin{center}
\includegraphics[width=0.7\textwidth]{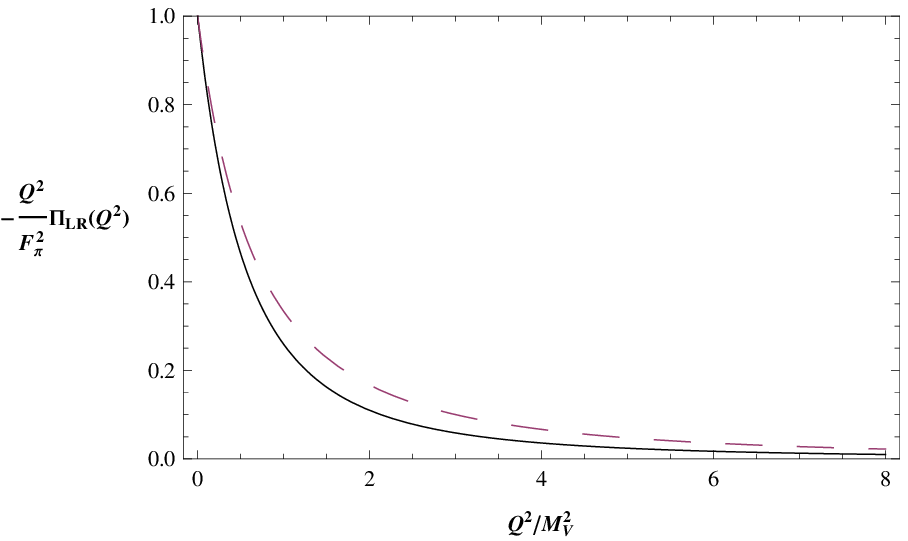}

\vspace*{0.25cm}
{\bf Fig.~5}
{\it\small $-\frac{Q^2}{M_V^2}\Pi_{\rm LR}(Q^2)$  for the Hurwith model (solid curve) and MHA (dashed curve).}

\end{center}
\end{figure}

From these considerations we conclude  that it is the Mellin transform of the spectral function (and its analytic continuation) which encodes all the  details of the underlying dynamics. One can also see that the determination of higher order condensates, the equivalent of the coefficients  in the high--$Q^2$ expansion in Eqs.~\rf{condmha} and \rf{condhm} is very sensitive to the nature of the hadronic spectrum. This is an important issue for phenomenology because, for example, the coefficient of $\cO(1/Q^6)$ in the expansion of $\Pi_{\rm LR}(Q^2)$ plays a crucial rôle in our understanding of direct CP--violation in $K$--decays, the so called $\epsilon'/\epsilon$ contribution~\cite{DG00}. Unfortunately, this $\cO(1/Q^6)$ term, as well as the higher order ones, appear to be very sensitive to the underlying dynamics. It is not surprising that the phenomenological determinations of these coefficients  found in the literature~\footnote{See e.g. refs.~\cite{GPP10,Metal11} and references therein.} differ so much. 

\section{\normalsize The Riemann Zeros and Sum Rules}\lbl{rzsr}
\setcounter{equation}{0}
\def\theequation{\arabic{section}.\arabic{equation}}

In the previous section we have  considered a model from Analytic Number Theory  to illustrate some physical features of Large--$\Nc$ QCD  which we would like to understand. Here I shall do the contrary, I will consider the physical framework of Dispersion Relations in Quantum Field Theory to discuss a well known problem in Mathematics. The problem has to do with the positions of the zeros of the Riemann zeta function defined by Eq.~\ref{Rie} and its analytic continuation~\footnote{For a nice elementary treatment of the Riemann zeta function see e.g. refs.~\cite{Havil03,Stopple03}.} which extends to all values of $s$ except at $s=1$ where it has a simple pole with residue $1$.

The interesting object for our purposes is the logarithmic derivative of the Riemann zeta function which, using the Euler product expression in Eq.~\ref{Rie}, can be written as a Dirichlet Series:
{\setl 
\bea\label{VonMD}
-\frac{\zeta'(s)}{\zeta(s)} & = & \sum_{{\rm primes}~p}\log(p)\sum_{k=1}^\infty p^{-ks}\nn \\
& = & {\underbrace{ \sum_{n=1}^\infty \Lambda(n)\  n^{-s}}_{\rm Dirichlet~Series}}\,,\quad \Ree(s)>1\,, 
\eea}

\noi
where the $\Lambda(n)$ ($n$ integer) are the so called {\it Von Mangoldt Amplitudes}:
\begin{equation}\lbl{VMamp}
\Lambda(n)  = \left\{\begin{array}{ll} \log(p)\,, & \mbox{if $n=p^k$ }\\ 
0\,, &  \mbox{otherwise} 
\end{array}\right. 
\end{equation}
A plot of the Von Mangoldt amplitudes for the first hundred integers is shown in Fig.~6.
\begin{figure}[h]

\begin{center}
\includegraphics[width=0.7\textwidth]{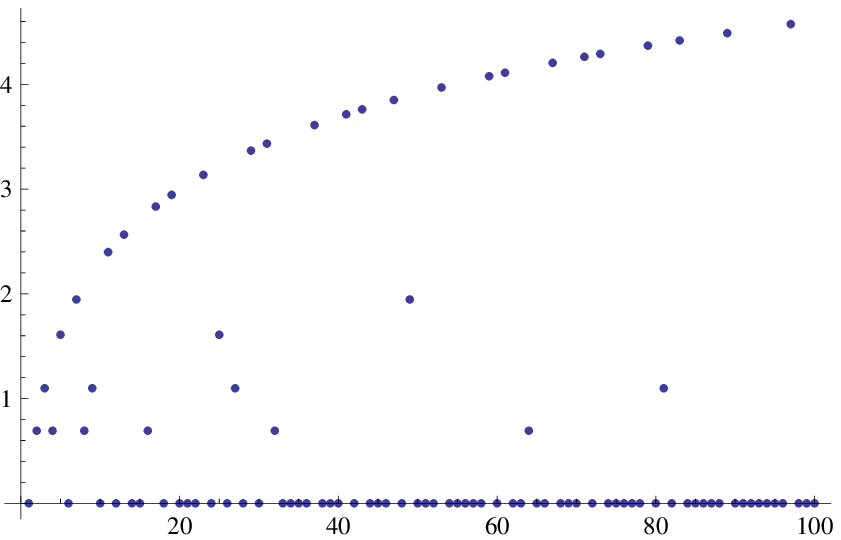}
\end{center}
\vspace*{0.10cm}
{\bf Fig.~6}
{\it\small  The Von Mangoldt values $\Lambda(n)$ for the first 100 integers. The upper curve corresponds to integer values coinciding with a prime number. The  {\it background} points correspond to integers which are more than a power of a prime number. They appear as successive horizontal points.
}

\end{figure}
In fact, there exists an explicit expression for the analytic continuation of the Dirichlet series in Eq.~\ref{VonMD}, which follows from Hadamard's product formula for $\zeta(s)$~\cite{Stopple03}:
\be\label{Hadamards} 
\Sigma_{\rm Von M}(s)  \equiv   -\frac{\zeta'(s)}{\zeta(s)}  =  \log\frac{1}{2\pi}+\frac{s}{s-1}
 +  \sum_{n=1}^\infty \frac{s}{2n(s+2n)}  -\sum_{\rho}\frac{s}{\rho(s-\rho)}\,.
\ee

\noi
The poles at $s=-2,\, -4\, -6\,\cdots$  correspond  to the {\it trivial zeros} of $\zeta(s)$ and the poles at $s=\rho$ to all the remaining zeros. The non--trivial zeros satisfy $0< \Ree (\rho)< 1\,,$ and  
because of the Symmetry Relation $s\rightarrow 1-s$, they must be located symmetrically relative to the vertical line $\Ree(s)=1/2\,,$ the so called {\it critical line}. The famous  
{\it Riemann Hypothesis} (RH) states that all the non--trivial zeros have $\Ree(s)=1/2\,.$ Numerically, all the non--trivial zeros which have been evaluated so far do indeed satisfy the RH. 

\subsection{\small Question}\lbl{question}
\noi 
{\it What are the properties of a Large--${\rm N}_c$ QCD--like Green's Function which has as a spectral function the one associated to the Von Mangoldt Dirichlet Series i.e., }
\begin{equation}\label{SVonMD}
\frac{1}{\pi}\Imm\Pi_{\rm VonM}(t)=\sum_{n=1}^{\infty}\Lambda(n) \ nM^2\ \delta(t-n M^2) \,,
\end{equation}
{\it with the ampltudes $\Lambda(n)$ given in Eq.~\rf{VMamp} and plotted in Fig.~6~?}
\\ \\ 
I wish to clarify the meaning of this question. From the point of view of QCD, this spectral function can only be considered, at best, as a toy model of large--${\rm N}_c$ (perhaps as a toy model of {\it duality violations}). What seems interesting to me, however, is the fact that such an abstract mathematical question as the RH can be phrased, as we shall see next, in terms of the language familiar to physicists working in Quantum Field Theory. 

Let us call $\Pi_{\rm VonM}(q^2)$ the two--point function which has as an imaginary part the {\it Von Mangoldt} spectral function in Eq.~\ref{SVonMD}.  The function $\Pi_{\rm VonM}(q^2)$  then obeys a dispersion relation modulo subtractions. The fact that the Van Mangoldt Dirichlet series in Eq.~\ref{VonMD} is defined for $\Ree(s)>1$ fixes the number of the required subtractions. They can be removed by taking two derivatives in the dispersion relation. This defines a function $\cP_{\rm VonM}(z)$ in the euclidean ($z=\frac{M^2}{Q^2}$), which is the analog of Eq.~\rf{displr}:
{\setl
\bea\label{VMSF}
\cP_{\rm VonM}(z)\equiv \frac{\partial^2}{(\partial Q^2)^2}\Pi_{\rm VonM}(Q^2) & = & 2\int_0^\infty dt\  \frac{Q^4}{(t+Q^2)^3}\ \frac{1}{\pi}\Imm\Pi_{\rm VonM}(t)\nn\\
 & = & 2\sum_{n=1}^\infty\Lambda(n)\frac{n z}{(1+n z)^3}\,,
\eea}

\noi 
with the corresponding Mellin--Barnes representation ($c=\text{Re}(s)\;\in\;]+1,+2[$)
\be
\cP_{\rm VonM}(z)  = 
 \frac{1}{2\pi i}\int\limits_{c-i\infty}^{c+i\infty} ds\ z^{-s} \ \Sigma_{\rm Von M}(s)\ \Gamma(s+1)\Gamma(2-s)\,, 
\ee
where $\Sigma_{\rm Von M}(s)$ is given in Eq.~\ref{Hadamards}. Everything is explicitly known in this representation and we can now apply the {\it Inverse Mapping Theorem} of ref.~\cite{FGD95} to compute the asymptotic behaviour of $\cP_{\rm VonM}(z)$.

The interesting expansion is the short distance expansion corresponding to large $Q^2$ (small--$z$). This expansion is governed by the singularities at the left of the fundamental strip i.e. $s\le 1$. The singularities at $s=-1,-2,-3,\cdots$ generated by the $\Gamma(s+1)$ factor in the integrand and by the poles in $\Sigma_{\rm Von M}(s)$ at $s=-2,-4,-6,\cdots$ corresponding to the trivial zeros of $\zeta(s)$ give rise to  odd powers of 
\begin{equation}
	\cO\left(\frac{M^2}{Q^2}\right)^{2n+1}\,,\quad n=0,1,2,3\cdots\,,
\end{equation}
as well as to even powers ($n=1,2,3\cdots$) of
\begin{equation}
\cO\left(\frac{M^2}{Q^2}\right)^{2n}\log\frac{Q^2}{M^2}\quad\annd\quad\cO\left(\frac{M^2}{Q^2}\right)^{2n}\,.
\end{equation}
These terms are rather analogous to the usual power terms which originate in the OPE in Quantum Field Theory. 
The leading singularity at $s=1$  gives rise to the leading  asymptotic behaviour for $Q^2$ large, which is:
\begin{equation}
\frac{1}{s-1}\Rightarrow \Gamma(2)\Gamma(1) z^{-1}= \frac{Q^2}{M^2}\,,	
\end{equation}
in fact rather similar to the leading behaviour of a QCD--like two--point function generated by a scalar current.

The  interesting terms are of course the ones generated by the next--to--leading singularities at $s=\rho$ in the $\Sigma_{\rm Von M}(s)$ function; i.e. the ones induced by the non--trivial zeros of the Riemann zeta function. They give rise to non--power terms:
\begin{equation}
-\frac{1}{s-\rho}\Rightarrow -\Gamma(\rho+1)\Gamma(2-\rho)z^{-\rho}\,,
\end{equation}
which appear in pairs of $\cO\left(\frac{Q^2}{M^2} \right)^{\vert\rho\vert}$ and 
$\cO\left(\frac{Q^2}{M^2} \right)^{1-\vert\rho\vert}$, modulated by an oscillating behaviour in $Q^2$. In the particular case where $\rho=1/2 \pm i\eta$, i.e. for the zeros satisfying the RH, these terms collapse to a unique non--power behaviour  of $\cO\left(\sqrt{\frac{Q^2}{M^2}}~\right)$~:
{\setl
\bea\label{1ntz}
\lefteqn{\frac{-1}{s-\left(\frac{1}{2} +i\eta\right)}+\frac{-1}{s-\left(\frac{1}{2} -i\eta\right)} \Rightarrow}  \nn \\
  &  & -\sqrt{\frac{Q^2}{M^2}}\frac{1 +4\eta^2}{2}\frac{\pi}{\cosh{\pi\eta}} \cos\left(\eta\log\frac{M^2}{Q^2}\right)\,, 
\eea}

\noi
modulated by oscillating $\cos\left(\eta\log\frac{M^2}{Q^2}\right)$ factors, one for each value of $\eta$ along the critical line of zeros, with amplitudes $\frac{1 +4\eta^2}{2}\frac{\pi}{\cosh{\pi\eta}}$ which decay exponentially for $\eta$ large. 

\vspace*{0.5cm} 

{\bf Conclusion:} {\it 
The Riemann Hypothesis is equivalent to the existence of a unique type of non--power terms of $\cO\left(\sqrt{\frac{Q^2}{M^2}}~\right)$ in the short distance expansion of the two--point function associated to the Von Mangoldt spectral function in Eq.~\ref{SVonMD}.}

\subsection{\small Quantum Mechanics Sum Rules}
\noi
It is perhaps helpful to discuss the previous considerations within a more general Statistical Mechanics Framework. 
The discussion applies to Hamiltonians $\cH$ with no explicit time
dependence. 

The probability transition amplitude in Quantum Mechanics is defined as
\begin{equation}
\langle q_{f},t_{f}\vert q_{i},t_{i}\rangle=
\langle q_{f}\vert e^{-i\cH (t_{f}-t_{i})}\vert q_{i}\rangle\,.
\end{equation}
The evolution in imaginary time leads to a Statistical Mechanics interpretation
 which is characterized by the {\it Partition Function}
\begin{equation}
\cZ=\mbox{Tr} \exp{-\beta\cH}\,.
\end{equation}
With $t_{f}-t_{i}=-i\beta$, we then have the spectral
representation:
\be
\langle q_{f}\vert e^{-\beta\cH}\vert q_{i}\rangle=\sum_{n}e^{-\beta
E_{n}}  \underbrace {\langle q_{f}\vert n\rangle\langle n\vert
q_{i}\rangle}_{
 \Psi_{n}(q_{f})\Psi^{*}_{n}(q_{i})}\,.
\ee
In particular, the $\lim \beta\rightarrow\infty$; i.e., $T\rightarrow 0$,  $[\beta=\frac{1}{kT}]$,
is governed by the {\it ground state} contribution:
\begin{equation}\lim_{\beta\rightarrow \infty}\langle q_{f}\vert e^{-\beta\cH}\vert
q_{i}\rangle\simeq e^{-\beta E_{0}}\Psi_{0}(q_{i})\Psi_{0}^{*}(q_{f})\,.
\end{equation}
In general:
\begin{equation}
E_{0}=\lim_{\beta\rightarrow +\infty} -\frac{1}{\beta}\log\Tr
e^{-\beta\cH}\,.
\end{equation}

The following quantity
\be\lbl{BB}
\cM(\beta)  \equiv  \langle q_{f}=0\vert e^{-\beta\cH}\vert q_{i}=0\rangle
  =  \sum_{n}\vert\Psi_{n}(0)\vert^{2} \exp[-\beta E_{n}]\,. 
\ee
is of special interest to us because it provides the Quantum Mechanics framework for the equivalent of the Quantum Field Theory  Sum Rules. 

The relevant Hamiltonian for our purposes is one which has levels
\begin{equation}
 E_n =n E_0\,,
\end{equation}
 and wave functions at the origin
\begin{equation}
\vert\Psi_{n}(0)\vert^{2}=\Lambda(n)\,,
\end{equation}
i.e. the Von Mangoldt amplitudes defined in Eq.~\rf{VMamp}.
Inserting this ansatz in Eq.~\rf{BB}, 
the corresponding Mellin--Barnes representation is then
\begin{equation}\label{VanMSM}
\cM_{\rm VonM}(\beta)=\frac{1}{2\pi i}\int\limits_{c-i\infty}^{c+i\infty} ds (\beta E_0)^{-s}\  \Sigma_{\rm VonM}(s)\ \Gamma(s)\,,
\end{equation}
with a {\it fundamental strip} defined now by the interval $c=\text{Re}(s)\;\in\;]+1,\infty[$  and $\Sigma_{\rm VonM}(s)$ the same expression as in Eq.~\ref{Hadamards}.
The {\it inverse mapping theorem} applied to this formula gives us the expansion at small $\beta$; i.e. the expansion at high temperature $\left(\beta=\frac{1}{kT}\right)$:

{\setl
\bea\lbl{betaexp}
\cM_{\rm VanM}(\beta) & \underset{{\beta\ra\  0}}{\thicksim} & \frac{1}{\beta E_0}
  -  \frac{1}{\sqrt{\beta E_0}}\sum_{\eta} \left[\Gamma(1/2 +i\eta)e^{i\eta\log(\beta E_0)}+\Gamma(1/2 -i\eta)e^{-i\eta\log(\beta E_0)}\right]\nn \\
 & +  & \log\frac{1}{4\pi} 
  +   \beta E_0\left(\log\left(8\pi \right)-\frac{1}{2}+\sum_{\rho}\frac{1}{\rho(1+\rho)}\right) \nn \\
  & +  & (\beta E_0)^2 \left[\frac{1}{2}\log(\beta E_0)-\frac{5}{12}-\frac{1}{2}\log(4\pi)+\frac{1}{2}\gamma_{\rm E}-\sum_{\rho}\frac{1}{\rho(2+\rho)}\right] \nn \\
  &  + & \cO\left[(\beta E_0)^3 \right]\,.
\eea}

\noi
In this expression  the leading behaviour in the first line is the one associated to the trivial singularity at $s=1$, while the second term in the first line gives the asymptotic behaviour reflected by the non--trivial zeros located at $s=1/2$. {\it The Riemann Hypothesis implies that these are the only possible non--power terms in the expansion.} 
A constant term as well as odd power terms in $(\beta E_0)$, with the leading one shown in the second line, are generated because of the $\Gamma(s)$ factor in Eq.~\ref{VanMSM} with coefficients which, except for the constant term,   depend on the location of the non--trivial zeros (the sum over $\rho$'s). The trivial singularities at $s=-2n$  generate  powers in $(\beta E_0)^{2n}$ (the leading one is shown in the third line) modulated by a $\log(\beta E_0)$ factor and powers of $(\beta E_0)^{2n}$. The latter are modulated by coefficients which also depend on the positions of the non--trivial zeros (the sum over $\rho$'s).

An interesting question which I have been investigating in collaboration with Josep Tar\'on  is: {\it what is the equivalent potential which produces a spectrum of discrete levels $E_n =nE_0$ with corresponding wave functions at the origin known in modulus $\vert\Psi_{n}(0)\vert^{2}=\Lambda(n)$ (i.e. the Von Mangoldt amplitudes defined in Eq.~\rf{VMamp})?} I hope to have an answer for Raymond's next birthday...
   
\newpage 
\begin{center}
{\normalsize\bf Acknowledgements.}
\end{center}

I felt very honored at giving a talk at the LAPP for the {\it Stora fest}. It is well known that Raymond's advice has been precious to many of us. He has had a great influence in keeping us honest! I am very happy to wish him:

\begin{center}
{\it Happy 80th Birthday Raymond with Many Returns of the Day}
\end{center}

\noi
I also wish to thank   M.~Knecht, S.~Peris, D.~Greynat, J.~Tar\'on and M.~Gonzalez-Alonso for many useful discussions on the topics presented here. 

This work has been partially supported by the EU RTN network FLAVIAnet [Contract No. MRTN-CT-2006-035482].


\vfill

\end{document}